# Deriving AC OPF Solutions via Proximal Policy Optimization for Secure and Economic Grid Operation

Yuhao Zhou, *Student Member*, *IEEE*, Bei Zhang, *Member*, *IEEE*, Chunlei Xu, Tu Lan, Ruisheng Diao, *Senior Member*, *IEEE*, Di Shi, *Senior Member*, *IEEE*, Zhiwei Wang, *Senior Member*, *IEEE*, and Wei-Jen Lee, *Fellow*, *IEEE*

*Abstract*—Optimal power flow (OPF) is a very fundamental but vital optimization problem in the power system, which aims at solving a specific objective function (ex.: generator costs) while maintaining the system in the stable and safe operations. In this paper, we adopted the start-of-the-art artificial intelligence (AI) techniques to train an agent aiming at solving the AC OPF problem, where the nonlinear power balance equations are considered. The modified IEEE-14 bus system were utilized to validate the proposed approach. The testing results showed a great potential of adopting AI techniques in the power system operations.

## I. INTRODUCTION

WITH the new resources and control strategies such as renewable energy, demand response, and energy storage devices being adopted into modern power grids, the future bulker power networks are increasingly complex, which accordingly introduces the requirement for fast control capability in the real-time grid operations to handle the randomness and fast-changing dynamics. Non-convex AC OPF is an important optimization problem that aims at solving a certain objective (e.g. operation costs) while maintaining the system in safe and stable operations. In recent decades, some important approaches were proposed to solve the AC OPF such as the primal-dual interior-point method [1]. Nonetheless, the results from the similar aforementioned method are usually locally optimal solutions. The convex relaxation method to tackle with AC OPF problem for radial networks via applying second-order cone programming (SOCP) was firstly proposed in [2]. In [3], solving AC OPF based on semidefinite programming (SDP) was proposed and the exactness of achieving the global optimum point was discussed. Nevertheless, as for all aforementioned methods, the iteration converging time is required to obtain the optimal solutions, which may be challenging and inadequate for the future power grids with the demand of a faster time scale control window.

Over the past few years, deep reinforcement learning (deep RL) has become increasingly popular to address challenging sequential decision-making problems and has been successfully applied in many areas such as games [4]-[5], robotics [6], autonomous driving [7], finance [8], and smart grid applications [9]. The optimal reactive power flow control based on deep RL was proposed in [10]. In [11], the deep Q network (DQN) was adopted to for smart building control. In [12]-[13], deep RL algorithms were adopted to perform the autonomous voltage control. In [14], the deep deterministic policy gradient algorithm was applied for the power system's load frequency control. Adaptive power system emergency control using deep RL was proposed in [15]. Applying deep RL in microgrid control were adopted in [16] and [17]. In [18], optimal management of the operation and maintenance of power grids based on RL was proposed, where the non-tabular RL algorithm adopting a neural network is designed and tested on the scaled-down power grid case study including renewable energy sources, controllable generators, maintenance delays and prognostics and health management devices. In addition, the deep RL was adopted to solve energy management problem in the field of energy Internet (EI) in [19]. In [20], the deep RL was utilized to handle voltage control and dynamic load shedding problem in a real power grid model. In [21], the deep RL was used to solve the EV charging scheduling problem for real-time control, which can adaptively learn the transition probability and does not require any system model information.

The main idea of this paper was inspired by aforementioned works to solve AC OPF problem, which aims at improving the solving speed compared with conventional optimization techniques. In order to increase the training speed and the performance of the trained agent, we applied neural network initialization process for the policy first, which is based on supervised learning technique. And then an agent is trained to handle AC OPF problem based on the state-of-the-art proximal policy optimization (PPO) algorithm [22], which belongs to one category of deep RL algorithms. In addition, in order to show the power of the deep RL technique, we modified the power grid branch flow limits to make the original AC OPF problem more difficult to solve. The training and testing platform in this paper is based on python and pypower, which is the python version of Matpower [23]. The modified IEEE 14 bus system is adopted to validate the effectiveness of the proposed method, which manifests a great promise of employing AI techniques in the future power system controls and operations.

## II. PRELIMINARIES AND BACKGROUND FOR PROXIMAL POLICY OPTIMIZATION ALGORITHM

In this section, it provides a brief introduction for deep RL and the algorithm of deep RL adopted in this paper.

### A. Brief introduction for Deep Reinforcement Learning

In general, RL could be depicted as a Markov decision process (MDP). The main idea of reinforcement learning (RL) is that an artificial intelligent (AI) agent may learn by interacting with its environment, which is similar to a biological agent [24]. An example of an agent interacting with its environment is shown in Fig. 1. The agent starts in a given initial state within its environment $s_0 \in S$ under a distribution of $p(s_0)$. At each time step $t$, the agent executes an action $a_t \in A$ according to its observed state $s_t$, while the environment emits an instantaneous/immediate reward $r_t \in R$ regarding the current action $a_t$ from the agent. And both the agent and the environment transit to the next state $s_{t+1}$ under the transition dynamics $T(s_{t+1}|s_t, a_t)$ that maps the distribution of states at time $t+1$ according to the current (state, action) pair at time $t$.

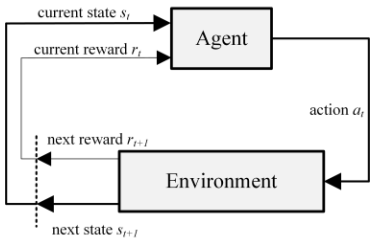

Fig. 1. The process of an agent interacting with the environment in RL.

A policy $\pi$ defines how an agent selects actions according to its observed states, and usually, it is categorized as deterministic or stochastic format:
- In deterministic case, the policy is defined as $\pi(s): S \to A$.
- In stochastic case, the policy is described by $\pi(s, a): S \to p(A = a | S)$, which maps from states to a probability distribution function over actions.

A value function represents a prediction of how good each state $V_\pi(s)$ or each (state, action) pair $Q^\pi(s, a)$ is. And advantage value function, $A^\pi(s, a)$, is introduced by applying the differences between $Q^\pi(s, a)$ and $V_\pi(s)$. If an agent only contains value functions or a policy function, then the RL algorithm is called a *value-based* or *policy-based* algorithm. What's more, the *actor-critic* employs both value functions and a policy function.

If the MDP is episodic, the states are reset after each episode length $T$, and a series of states, actions and rewards within an episode in a sequence constitute *trajectories* or *rollouts* of the policy, which are regarded as the collected past experiences. Every trajectory of a policy returns an accumulated reward shown in equation (1):

$$R = \sum_{t=0}^{T-1} \gamma^t r_{t+1} \qquad (1)$$

where $\gamma \in [0, 1]$ is a discount factor and a lower value of $\gamma$ is more emphasis on the instantaneous rewards.

The goal of RL is to find an optimal policy $\pi^*$ that maximizes the expected return $\mathbb{E}$ for all states shown as follows.

$$\pi^* = \arg\max_\pi \mathbb{E}(R | \pi) \qquad (2)$$

By combing with deep neural networks (NN) as approximations for value and policy functions to tackle high dimensional state and action spaces, deep RL has become a powerful approach to handle real-world problems.

### B. Proximal Policy Optimization with Clipped Surrogate Loss

In the reinforcement learning field, the policies could be directly optimized by policy gradient algorithms during the training process. Unlike the value-based deep RL algorithms such as deep Q network (DQN) to discretize the action spaces, not only policy gradient algorithms are more suitable and effective in high dimensional and/or continuous action spaces with better convergence properties, but also can learn stochastic policies. A lot of progress has been made recently based on policy gradient algorithms such as deep deterministic policy gradient (DDPG) [25], asynchronous advantage actor-critic (A3C) [26], trust region policy optimization (TRPO) [27], PPO [15] and so on.

Compared with other state-of-the-art policy gradient approaches, PPO performs comparable or better than others, which makes it become the default deep RL algorithm in OpenAI. Therefore, in this paper, we applied PPO to achieve the optimal policy for the trained agent.

PPO is in the *actor-critic* format, which contains an actor NN and a critic NN. PPO takes advantage of the data efficiency, reliable and stable performance of TRPO, but avoids the computational complexity of second-order optimization by applying only first-order optimizers like the Gradient Descent method to achieve the optimal objectives. In the trust region, KL-divergence, which describes the differences between two data distributions, is applied as the maximum step size for the agent to explore. Unlike TRPO using KL-divergence constraint, PPO considers modifying the objective function by applying a clipped surrogate objective function shown as follows.

$$L(\theta) = \hat{\mathbb{E}}_t[\min(r_t(\theta) A^{\pi_{\theta_{old}}}(s_t, a_t), \text{clip}(r_t(\theta), 1-\varepsilon, 1+\varepsilon) A^{\pi_{\theta_{old}}}(s_t, a_t))]$$

where $r_t(\theta) = \dfrac{\pi_\theta(a_t / s_t)}{\pi_{\theta_{old}}(a_t / s_t)}$ \qquad (3)

In equation (3), $\theta$ is the actor NN's parameters, which learns an optimal policy $\pi_\theta(a_t|s_t)$. In addition, the ratio between the updated new policy and old policy, $r_t(\theta)$, measures how the difference between the two policies. $\varepsilon$ is a newly introduced hyper-parameter to determine the clipping range. If the $r_t(\theta)$ falls outside the range $(1-\varepsilon)$ and $(1+\varepsilon)$, the advantage function will be clipped, which not only could guarantee the stability of this algorithm but also make it simple to perform optimization via first-order methods.

In the PPO algorithm, generalized advantage estimation (GAE) function is employed to calculate advantage value [28], where $V(s_t)$ is the state value that comes from the critic NN.

$$\begin{cases} \hat{A}_t^{GAE(\gamma,\lambda)} = (1-\lambda)(\hat{A}_t^{(1)} + \lambda \hat{A}_t^{(2)} + \lambda^2 \hat{A}_t^{(3)} + ...) \\ \hat{A}_t^{(k)} = \sum_{l=0}^{k-1} \gamma^l \delta_{t+l}^V = -V(s_t) + r_t + \gamma r_{t+1} + ... + \gamma^{k-1} r_{t+k-1} + \gamma^k V(s_{t+k}) \end{cases} \quad (4)$$

What's more, the policy is parameterized as a conditionally Gaussian policy $\pi_\theta \sim \mathcal{N}(\mu_\theta(s), \Sigma_{\pi\theta})$ in PPO algorithm, where $\mu_\theta(s)$ is the output of the actor NN, and $\Sigma_{\pi\theta}$ is the covariance matrix of Gaussian policies.

The objective function for the critic NN is shown as follows.

$$\min_{\varphi} \sum_{(s_t, r_{dis\_sum}) \in \mathcal{D}_{batch}} \frac{1}{N} \| r_{dis\_sum} - V_\varphi(s_t) \|_2^2 \quad (5)$$

where $r_{dis\_sum}$ is discounted accumulated reward, and $\varphi$ is the critic NN parameters.

## III. THE PROCEDURES OF THE PROPOSED TRAINING PROCESS FOR PPO BASED AC OPF

The proposed methodology of developing the training process to deal with the AC OPF problem could be outlined as follows. Firstly, an appropriate power grid environment used for deep RL should be established, where appropriate states, actions, and a reward function would be determined. Then, the NN initialization process is applied for establishing the initialized policy from mimicking the AC OPF solver "behaviors" of generator settings, which could be seen as the initial status of the agent instead of using random initial parameters for the policy neural network. Finally, the PPO is applied to further train the agent such that better performance could be achieved.

A brief description of the agent interacting with the power grid environment based on Fig. 1 is explained as follows.

The "*episodes*" represent the real-time power system operating conditions collected from SCADA or PMU such as different load variations and generator dispatches, which could be used to form training and testing datasets. However, the changing of topology, contingencies and transient responses of the system are not considered yet in this paper. The "*maximum episode length*" could be interpreted as the maximum steps of actions that an agent is allowed to take within one episode.

### A. Generic Grid Environment used for Deep RL

In this paper, the environment is developed by mimicking the OpenAI Gym environments, which are the benchmark systems used for deep RL studies.

Under the episode *i*, which is procured from the current operating condition including the information of current loads and generator settings, the function env.reset() is applied to obtain the initial state $s_t$ by running Newton-Raphson power flow (PF) solver (enforced with generator Q-limits). Here "*env*" stands for the class object of the grid environment. Then according to the initial state $s_t$, the agent gives an action $a_t$ to tune the generators. Subsequently, the function env.step($a_t$) is applied in the environment, where the next state $s_{t+1}$ and the immediate reward $r_t$ for the action $a_t$ are returned to the agent as one past experience.

Usually, the function env.step($a_t$) like in OpenAI gym would also provide an ending condition flag called "*done*" for the episode, which indicates that the agent has accomplished the pre-set goal for the control objects. However, it's very tricky to decide "*done*" condition for solving the OPF problem because it is unknown that when and whether the agent's actions have made the system to achieve the optimal status regarding the current load and generator settings. On the other hand, load variations and dissimilar generator settings may contribute to different optimal status, which also makes it challenging to design a sensible reward function. Therefore, the "*done*" flag setting of one episode is determined in this paper as follows: if the power flow solver diverges, this episode is ended (the "*done*" flag is 1) and it will go to the next episode, otherwise the environment updates the current state and continues interacting with the agent until the maximum episode length *T* is reached or power flow solver diverges (the "*done*" flag is 1).

What's more, a good reward function would pinpoint good guidance and may lead to faster learning speed for the agent. The detailed design for the reward function in this paper is shown as follows.

$$\text{reward} = \begin{cases} -5000, & \text{if power flow diverges} \\ r_{pgen\_v} + r_{vbus\_v} + r_{line\_v}, & \text{if constraints violations are happened} \\ k \times cost_{generators} + b + z, & \text{if no violations and power flow converges} \end{cases} \quad (6)$$

The $r_{pgen\_v}$, $r_{vbus\_v}$, and $R_{br\_v}$ correspond to the negative penalty rewards if the actions violate: (1) the active power outputs of all generators inequality constraints, (2) the voltage magnitude limits for every bus in the system, (3) the branch line flow limits for every transmission line in the system including two directions, respectively. In order to reflect the fact that optimal actions under different operating conditions should have similar rewards even though the objective costs are totally different, in this paper, the coefficients *k* and *b* are applied to map different optimal actions within a similar range of rewards, for example, [0, 500] points, and the coefficient *z* is considered as one correction item to compensate the previous mapping errors.

### B. States and Action Space

As for a power grid system containing *m* buses and *s* generators, in this paper, the state, which is the input for the policy and value function represented by neural networks, includes the following variables: the load's active power in the power grid $P_{da}$, and the load's reactive power $Q_{da}$ in the power grid for every bus *a*, every generator *t*'s initial active power setting $P_{gt}$ and its voltage setting $V_{gt}$, and they are concatenated shown as follows.

$$\text{state} = [P_{d1},...,P_{dm}, Q_{d1},...,Q_{dm}, P_{g1},...,P_{gs}, V_{g1},...,V_{gs}] \quad (7)$$

As for the action space, it consists of the tuning step size of all generator settings in a row vector format similarly shown as follows.

$$\text{action} = [\Delta P_{g1},...,\Delta P_{gs}, \Delta V_{g1},...,\Delta V_{gs}] \quad (8)$$

### C. Neural Network Structures in PPO Training

The actor NN structure in PPO are shown in Fig. 2. The input for the actor NN is shown in equation (7), where the input shape is [*N*, 2*(n+t)*] with training batch size *N*. As for the structure of the actor NN in this paper, only the first *2\*n* columns of data are used as the input for the neural network, and there are *m* hidden layers with *Relu* activation function. The outputs of the output layer could be seen as the optimal settings in per unit value for all generators in the system with the shape of *[N, 2*t]*, which is reasonable to apply *Sigmoid* activation function in the output layer. Meanwhile, the output of the actor NN $\mu_\theta(s)$ provides the mean values of the conditionally Gaussian policy $\pi_\theta$.

As for the structure of the critic NN, similarly the input shape is *[N, 2*(n+t)]*, nevertheless all columns of data are connected with the input layer, and there are *y* hidden layers with *Relu* activation function. The output shape of the critic NN is [*N*, *1*], where only one neuron is set in the output layer.

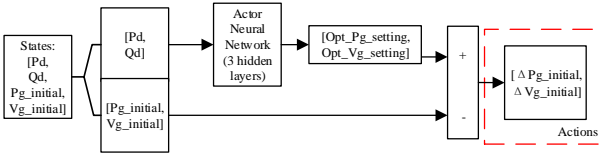

Fig. 2. The NN structure for actor policy in PPO.

### D. Neural Network training Process

Learning from scratch may need a much higher sample complexity, which leads to longer training time, and it may fail when the dimensionality of the action space is very high. Therefore, in this paper, we adopted NN initialization process before the applying PPO algorithm to further train the network.

The "labels" for the actions during the initialization process could be accumulated by running AC OPF solver offline. Afterward, the inputs, which are the states and the the "expert" actions are utilized to initialize the actor NN as a regression problem by applying the supervised learning algorithm, where mean square error between the outputs of the actor NN $\mu_\theta(s)$ parameterized by $\theta$ and the "labels" $\hat{a}$ including optimal active power actions and voltage actions. By applying the Stochastic Gradient Descent optimizer, the actor NN could be trained to clone the optimal generator settings from OPF solver results.

Then by applying the PPO algorithm in [22], the neural network could be further trained and the advantages of deep RL is shown, which would be further discussed in Section IV.

## IV. CASE STUDY

The proposed approach applying PPO for training an agent to solve the AC OPF problem was validated in this section by adopting the IEEE 14 bus system. The simulation platform is based on TensorFlow and PYPOWER, which provides the platform for NN training, the Newton-Raphson power flow (PF) solver and PYPOWER interior-point AC OPF solver (PIPS), respectively.

The data generating process is illustrated as follows. Compared with default load information, the system loads are perturbed randomly between [0.6, 1.4] times. And the system generators' initial settings are similarly perturbed randomly between [**P**$_{gmin}$, **P**$_{gmax}$] and [**V**$_{gmin}$, **V**$_{gmax}$], where **P**$_g$ and **V**$_g$ denote the vectors for every generator's active power setting and voltage setting. In addition, the embedded PIPS is adopted to generate the optimal action labels, which would be used in NN initialization and to indicate whether the OPF problem is feasible. Accordingly, the data for both the IEEE 14 bus system with only feasible OPF scenarios are accumulated, which is then divided into a training dataset, testing dataset respectively. Moreover, due to the random perturbations for initial generator settings and the fact that only the equality constraint is considered in the Newton-Raphson PF solver, though it is converged in PYPOWER environment, there are still possible "abnormal" situations such as the results of generators' active power outputs reside outside the limits. What's more, only the training dataset is used for NN initialization and PPO training, while the testing dataset is applied to verify the training results. Additionally, during the testing process, the outputs of the actor NN are regarded as the optimal actions directly instead of sampling from the trained stochastic policy, and the episode length is set as 5.

The dimensions of state space and action space are 38 and 10 respectively in this system. The line 4-5 line flow limit is replaced by 32 MVA, which makes it more difficult to handle with the original load information.

There is 55000 data that was generated as the training dataset, and 17364 data that is considered as the testing dataset. Firstly, the NN initialization process was conducted, where the data is shuffled first and 99% of the training data was applied to finish this process. Accordingly, the left 1% of the training dataset is considered as the testing dataset for this NN initialization process. And the average loss value for the testing dataset is shown in Table I.

TABLE I
THE AVERAGE LOSS VALUES FOR THE NN INITIALIZATION PROCESS

| Testing system | Average MSE value for Generators' active power setting (MW) | Average MSE value for Generators' active power setting (p.u.) |
|---|---|---|
| IEEE 14-bus system | 0.435 | 2.6e-7 |

From Table I, it is seen that the losses are very small, which validates the structure of actor NN. Then we applied this actor NN from the initialization process applying supervised learning technique to perform the AC OPF task testing process regarding the testing dataset, and the results are shown in Fig. 3. Because the objective function (total generator's costs) may be smaller than the PIPS solver results due to the violations, in order to reflect the violation status instead of only focusing on the objective function values, the $20000 or $30000 is manually set as cost values for constraints violation or divergence scenarios, respectively.

From Fig. 3, though the training loss values shown in the Table I are very small, 58.89% data in the testing dataset still violates the system constraints. And from the second row of the Fig. 3, it's seen that most of the violations are line flow limits due to the modification of line flow limits for the line 4-5's in the system.

In order to further show the advantages of deep RL, the initialized policy comes from the result that only 18% data in training data set is employed. Then this initialized agent is trained by applying PPO algorithm. Afterwards, in the same way, we applied this trained NN to perform the AC OPF task regarding the same testing dataset. The corresponding results are shown in Table II.

From Table II, after adopting PPO training with initialization, 100% data in the testing dataset could achieve optimal status, and the optimal costs from the agent are very close to the results from PIPS, where the maximum cost deviation in percentage from the trained agent is around 1.75% higher, which is verified that the proposed method is valid in a small power system such as the IEEE 14 bus system. And with the implementation of deep RL algorithm to train the neural network, the success rate of performing AC OPF task is increased significantly.

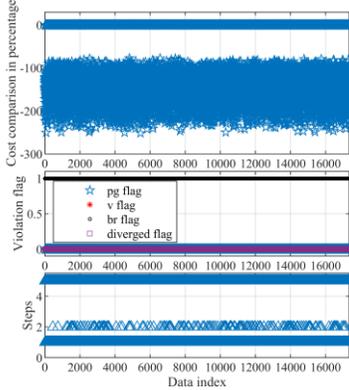

Fig. 3. AC OPF task results of testing dataset from NN initialization process.

TABLE II
THE COMPARISON RESULTS FOR THE TRAINED AGENT WITH APPLYING PPO AND WITHOUT APPLYING PPO

| Trained agent | success rate |
|---|---|
| NN trained by supervised learning technique | 41.11% |
| Initialized NN with PPO training | 100% |

## V. CONCLUSION

In this paper, we proposed a framework to solve the OPF problem by applying the state-of-the-art deep RL algorithm. The IEEE 14 bus system was utilized to verify the proposed approach. With the help of implementing the deep reinforcement learning, the success rate of performing OPF task could be significantly improved, which validated the effectiveness of the proposed method and manifests a great promise of adopting AI techniques in the power system operations and controls.

## VI. REFERENCES

[1] R. A. Jabr, "A primal-dual interior-point method to solve the optimal power flow dispatching problem," *Optimization and Eng.*, vol. 4, no. 4, pp. 309–336, Dec. 2003.
[2] R.A. Jabr, "Radial distribution load flow using conic programming," *IEEE Trans. Power Syst.*, vol. 21, no. 3, pp. 1458–1459, Aug. 2006.
[3] J. Lavaei, and S. H. Low, "Zero duality gap in optimal power flow," *IEEE Trans. Power System*, vol. 27, no. 1, pp. 92-107, Feb. 2012.
[4] V. Mnih, K. Kavukcuoglu, D. Silver, A. Graves, I. Antonoglou, D. Wierstra, and M. Riedmiller. "Playing Atari with deep reinforcement learning". In: arXiv preprint arXiv: 1312.5602 (2013).
[5] D. Silver, A. Huang, C. J. Maddison, A. Guez, …, and S. Dieleman, "Mastering the game of Go with deep neural networks and tree search," *Nature*, vol. 529, no. 7584, pp. 484-489, Jan. 2016.
[6] A. Nagabandi, G. Kahn, R. S. Fearing, and S. Levine. "Neural network dynamics for model-based deep reinforcement learning with model-free fine-tuning". In: arXiv preprint arXiv: 1708.02596 (2017).
[7] X. Pan, Y. You, Z. Wang, and C. Lu. "Virtual to real reinforcement learning for autonomous driving". In: arXiv preprint arXiv: 1704.03952 (2017).
[8] Y. Deng, F. Bao, Y. Kong, Z. Ren, and Q. Dai, "Deep direct reinforcement learning for financial signal representation and trading," *IEEE Trans. Neural Networks and Learning Systems*, vol. 28, no. 3, pp. 653-664, March 2017.
[9] D. Zhang, X. Han, and C. Deng, "Review on the research and practice of deep reinforcement learning," *CSEE Journal of Power and Energy Systems*, vol. 4, no. 3, pp. 362–370, Sep. 2018.
[10] Y. Xu, W. Zhang, W. Liu, and F. Ferrese, "Multiagent-based reinforcement learning for optimal reactive power dispatch," *IEEE Trans. Systems, man, and cybernetics, part C (Applications and Reviews)*, vol. 42, no. 6, pp. 1742-1751, Nov. 2012.
[11] E. Mocanu, D. C. Mocanu, P. H. Nguyen, A. Liotta, M. E. Webber, M. Gibescu, and J. G. Slootweg, "On-line building energy optimization using deep reinforcement learning," *IEEE Trans. Smart Grid*, vol. 10, no. 4, pp. 3698–3708, July 2019.
[12] R. Diao, Z. Wang, D. Shi, Q. Chang, J. Duan, and X. Zhang. "Autonomous voltage control for grid operation using deep reinforcement learning". In: arXiv preprint arXiv: 1904.10597 (2019).
[13] J. Duan, D. Shi, R. Diao, H. Li, Z. Wang, B. Zhang, D. Bian, and Z. Yi, "Deep-reinforcement-learning-based autonomous voltage control for power grid operations," *IEEE Trans. Power System*, Early Access, Sep. 2019.
[14] Z. Yan, and Y. Xu, "Data-driven load frequency control for stochastic power systems: a deep reinforcement learning method with continuous action search," *IEEE Trans. Power System*, vol. 34, no. 2, pp. 1653–1656, March 2019.
[15] Q. Huang, R. Huang, W. Hao, J. Tan, R. Fan, and Z. Huang, "Adaptive power system emergency control using deep reinforcement learning," *IEEE Trans. Smart Grid*, vol. 11, no. 2, pp. 1171–1182, March 2020.
[16] Y. Ji, J. Wang, J. Xu, X. Fang, and H. Zhang, "Real-time energy management of a microgrid using deep reinforcement learning," *Energies*, vol. 12, no. 2, June 2019.
[17] François-Lavet, V., Taralla, D., Ernst, D. and Fonteneau, R., 2016. Deep reinforcement learning solutions for energy microgrids management. In *European Workshop on Reinforcement Learning (EWRL 2016)*.
[18] R. Rocchetta, L. Bellani, M. Compare, E. Zio and E. Patelli, "A reinforcement learning framework for optimal operation and maintenance of power grids," *Applied energy*, *241*, pp.291-301, 2019
[19] H. Hua, Y. Qin, C. Hao, and J. Cao, "Optimal energy management strategies for energy Internet via deep reinforcement learning approach," *Applied energy*, *239*, pp.598-609, 2019
[20] J. Zhang, C. Lu, J. Si, J. Song and Y. Su, July. Deep reinforcement learning for short-term voltage control by dynamic load shedding in China Southern power grid. In *IEEE 2018 International joint conference on neural networks (IJCNN)* (pp. 1-8).
[21] Z. Wan, H. Li, H. He, and D. Prokhorov, "Model-free real-time EV charging scheduling based on deep reinforcement learning," *IEEE Trans. Smart Grid*, vol. 10, no. 5, pp. 5246–5257, Sept. 2019.
[22] J. Schulman, F. Wolski, P. Dhariwal, A. Radford, and O. Klimov. "Proximal policy optimization algorithms". In: arXiv preprint arXiv: 1707.06347 (2017).
[23] R. D. Zimmerman, C. E. Sanchez, and R. J. Thomas, "MATPOWER: steady-state operations, planning, and analysis tools for power systems research and education," *IEEE Transactions on Power Systems*, vol. 26, no. 1, pp. 12-19, 2011.
[24] V. Francois-Lavet, P. Henderson, R. Islam, M. G. Bellemare, and J. Pineau. "An introduction to deep reinforcement learning". In: arXiv preprint arXiv: 1811.12560 (2018).
[25] T. P. Lillicrap, J. J. Hunt, A. Pritzel, N. Heess, T. Erez, Y. Tassa, D. Silver, and D. Wierstra. "Continous control with deep reinforcement learning". In: arXiv preprint arXiv: 1509.02971 (2015).
[26] V. Minh, A. P. Badia, M. Mirza, A. Graves, T. P. Lillicrap, T. Harley, D. Silver, and K. Kavukcuoglu. "Asynchronous methods for deep reinforcement learning". In: arXiv preprint arXiv: 1602.01783 (2016).
[27] J. Schulman, S. Levine, P. Moritz, M. I. Jordan, and P. Abbeel. "Trust region policy optimization". In: arXiv preprint arXiv: 1502.05477 (2015).
[28] J. Schulman, P. Moritz, S. Levine, M. Jordan, and P. Abbeel. "High-dimensional continuous control using generalized advantage estimation". In: arXiv preprint arXiv: 1506.02438 (2015).